\documentclass[prx,twocolumn,floatfix,superscriptaddress,notitlepage,longbibliography]{revtex4-2}

\usepackage{changes}
\setdeletedmarkup{\color{blue}\sout{#1}}
\usepackage{cancel}

\usepackage[dvipsnames]{xcolor}
\usepackage[most]{tcolorbox}

\usepackage[normalem]{ulem}
\usepackage{graphicx, color, graphpap}
\usepackage{mathrsfs} 
\usepackage{enumitem}
\usepackage{amssymb}
\usepackage{amsthm}
\usepackage{multirow}
\usepackage[colorlinks=true,citecolor=blue,linkcolor=blue,urlcolor=blue]{hyperref}
\usepackage[T1]{fontenc}
\usepackage{verbatim}
\usepackage{mathtools}
\usepackage{titlesec}

\long\def\ca#1\cb{} 

\newcommand{\abs}[2][]{#1| #2 #1|}

\newcommand{\ketbra}[2]{| \hspace{1pt} #1 \rangle \langle #2 \hspace{1pt} |}

\newcommand{\norm}[2][]{#1| \! #1| #2 #1| \! #1|}

\newcommand{\ket}[1]{|#1\rangle}               
\newcommand{\bra}[1]{\langle #1|}              
\newcommand{\dya}[1]{\ket{#1}\!\bra{#1}}

\newcommand{\Tr}{{\rm Tr}}

\renewcommand{\geq}{\geqslant}
\renewcommand{\leq}{\leqslant}

\newcommand{\ad}{^\dagger}

\newcommand*{\id}{\openone}

\newcommand{\tot}{\text{tot}}
\newcommand{\eq}{\text{eq}}

{}
{}

\begin{document}
\title{Information Processing in Quantum Thermodynamic Systems:\\ an Autonomous Hamiltonian Approach }

\author{Shou-I Tang}
\affiliation{Department of Physics, University of Massachusetts, Boston, Massachusetts 02125, USA}

\author{Emery Doucet}
\affiliation{Department of Physics, University of Massachusetts, Boston, Massachusetts 02125, USA}

\author{Akram Touil}
\affiliation{Theoretical Division, Los Alamos National Laboratory, Los Alamos, NM 87545, USA}

\author{Sebastian Deffner}
\affiliation{Department of Physics, University of Maryland, Baltimore County, Baltimore, MD 21250, USA}
\affiliation{Quantum Science Institute, University of Maryland, Baltimore County, Baltimore, MD 21250, USA}
\affiliation{National Quantum Laboratory, College Park, MD 20740, USA}

\author{Akira Sone}
\email{akira.sone@umb.edu}
\affiliation{Department of Physics, University of Massachusetts, Boston, Massachusetts 02125, USA}
\affiliation{The NSF AI Institute for Artificial Intelligence and Fundamental Interactions}
	
\begin{abstract}
Extending the quantum formulation of [Phys. Rev. X \textbf{3}, 041003 (2013)] to a more general setting for studying the thermodynamics of information processing including initial correlations, we generalize the second law of thermodynamics to account for information processing in such autonomous systems. We consider a composite quantum system consisting of a principal system, heat bath, memory, and work source, and adopt an autonomous Hamiltonian framework. We present the following three main results. We first derive constraints on the total Hamiltonian that ensure the work source to act as a catalyst preserving its original randomness, given that the total unitary evolution must have a unitary partial transpose. 
Second, we generalize the quantum speed limit for the joint dynamics of system and memory to the \emph{quantum thermodynamic speed limit}, from which we obtain a dynamical version of Landauer’s bound. Finally, we also interpret this quantum thermodynamic speed limit in the context of quantum hypothesis testing.
\end{abstract}

\maketitle
\section{Introduction}
\label{sec:intro}

With the rapid development of  viable technologies for quantum information processing (QIP), a wide range of potential applications are envisioned, many of which are expected to have a transformative impact~\cite{Nielsen}. Examples include quantum machine learning~\cite{schutt2020machine,biamonte2017quantum,jerbi2023quantum,torlai2020machine,schuld2021machine,gao2018quantum,gao2022enhancing,sone2024quantum}, which leverages quantum resources to enhance data analysis and pattern recognition; quantum simulation~\cite{georgescu2014quantum,maskara2025programmable,puig2025variational,halimeh2025cold,shapira2025programmable}, which enables the exploration of complex quantum many-body systems beyond the reach of classical computation; and quantum metrology~\cite{degen2017quantum, giovannetti2006quantum, giovannetti2011advances,cerezo2021sub,barbieri2022optical,zhou2020quantum,sone2020generalized,beckey2022variational,zhang2021distributed}, which seeks precision surpassing classical limits.

While current platforms operate in the noisy intermediate-scale quantum (NISQ) era~\cite{preskill2018quantum,cerezo2021barren,cerezo2021variational,wang2021noise,preskill2025beyond} and provide valuable insights into algorithm design \cite{bharti2022noisy,Domino2023entropy,Smierzchalski2024SR,Robertson2025entropy,Domino2025SR,Robertson2025} {, error mitigation \cite{Roffe2019CP,Cavalcante2025EPL}, and thermodynamic resources \cite{DeffnerBook19,Aifer2023PRXQ,Smierzchalski2024SR_2,Stevens2025QST}}, realizing  applications with genuine quantum advantage ultimately requires scalable and fault-tolerant quantum  hardware~\cite{gottesman1998theory,shor1996fault,sunami2025scalable,bluvstein2025architectural,katabarwa2024early}. A key step toward this goal is developing a deep understanding of the fundamental physical constraints of information processing in quantum systems. Two classes of constraints are of particular significance. The first is energetic in nature and is captured by the quantum Landauer principle~\cite{landauer1961irreversibility,goold2015nonequilibrium, van2022finite, reeb2014improved, timpanaro2020landauer,Deffner2021EPL,chattopadhyay2025landauer}, which sets a fundamental thermodynamic bound on the minimal energy required for information erasure and relevant information processing in quantum systems. The second is dynamical, expressed through quantum speed limits (QSL)~\cite{deffner2017quantum, nishiyama2025speed, aifer2022quantum, campaioli2019tight, frey2016quantum,deffner2020quantum,deffnerquantum2013, deffner2017geometric,mohan2022quantum}, which set ultimate bounds on how fast quantum states can evolve under given physical resources. These constraints not only impose limitations but also serve as guiding principles for the design of efficient protocols \cite{Aifer2022NJP} and architectures. Therefore, by exploring the interplay between thermodynamic and dynamical constraints crucial insight can be obtained into how to realize fault-tolerant quantum information processors.

Quantum thermodynamics~\cite{DeffnerBook19,Binder19,Anders16,Goold16,Auffeves2022energy,campbell2025roadmap} has been developed to explore the behavior of thermodynamic laws at the nanoscale during finite-time  {evolution} by considering the role of quantum resources, such as quantum coherence~\cite{gour2022role, bernardo2020unraveling, hammam2022exploiting, kwon2018clock,sone2021quantum,Kwon2025}, entanglement~\cite{oppenheim2002thermodynamical,touil2021ergotropy,sone2025no,sone2023exchange}, quantum fluctuations~\cite{jarzynski1997nonequilibrium,crooks1999entropy,tasaki2000jarzynski,kurchan2000quantum,sone2023jarzynski, kacper2024quantum,maeda2023detailed,Manzano18,Campisi09,sone2025thermodynamic}, and quantum nonlinearity \cite{Deffner2025QST,Deffner2025GPQ}.  Therefore, quantum thermodynamics is the natural framework to investigate the interplay between  energetic resources and dynamical constraints, encoded in the quantum Landauer principle and the QSL, respectively.  Namely, analyzing the thermodynamic cost of information processing in quantum systems  quantifies how the minimal energetic resources depend on the quantum operation time. At the same time, QSL  relates the rate of computation to the available energy and the dynamics of the system. Therefore, quantum thermodynamics allows one to study these connections in a systematic way and may assist in developing optimized strategies to approach both the energetic and temporal limits in practical quantum devices.

 Reference~\cite{deffner2013information} developed a comprehensive framework for the thermodynamics of  classical information processing, where devices, a heat bath, a work source, and a memory reservoir are all explicitly modeled and evolve autonomously under a time-independent Hamiltonian. Specifically, the work source is assumed to maintain its initial randomness throughout the evolution, and the energy of the memory reservoir is considered to stay constant during the process.  Key results are generalizations of the Kelvin-Planck, Clausius, and Carnot statements  in the presence of information gathering, establishing that Shannon entropy should be regarded on equal footing with Clausius entropy in  formulations of the second law. This perspective  elucidates, for instance, that extracting work from a single heat bath or efficiencies beyond Carnot  are perfectly thermodynamically sound, provided  that the associated entropy reduction is balanced by information storage  in the memory.  The distinguishing feature of this framework is an autonomous, fully inclusive treatment, in which the memory, also called \emph{information reservoir} plays an essential role.

In the present paper, we extend the framework of Ref.~\cite{deffner2013information} to the quantum domain.
In  our quantum setting, Liouville dynamics is replaced by unitary evolution under the autonomous (time-independent) total Hamiltonian, Shannon entropy by von Neumann entropy, and classical ensemble averages by quantum expectation values with respect to density operators. We emphasize autonomous Hamiltonians  since they provide a closed, self-contained description of all subsystems, ensuring that energy and entropy exchanges among the subsystems arise solely from internal interactions, without the need for externally imposed controls.  Consequently, thermodynamic laws emerge naturally and self-consistently from the underlying unitary dynamics. Within this framework, we determine the conditions that the total Hamiltonian must satisfy to be thermodynamically consistent, derive a quantum version of Landauer’s principle to capture energetic constraints, and investigate the effective QSL for the composite of the principal system and memory. We further examine the relationship between the Landauer bound and the effective QSL, elucidating their operational significance in the context of quantum hypothesis testing.

This paper is organized as follows. Extending upon Ref.~\cite{deffner2013information}, we present our setup in Sec.~\ref{sec:setup}. In Sec.~\ref{sec:unitaryAndhamiltonian}, we examine the unitary operator governing the total time evolution and Hamiltonian, showing that its partial transpose must also be unitary and that the time evolution must be compatible with the state of the work source~\cite{Lie21}. Based on these, we formulate the first and second laws of thermodynamics in Sec.~\ref{sec:laws}.
In Sec.~\ref{sec:QTSL}, we introduce the quantum thermodynamic speed limit characterized and establish its connection to the quantum Landauer principle.
Finally, in Sec.~\ref{sec:conc}, we demonstrate that the quantum thermodynamic speed limit determines the scaling ratio in quantum hypothesis testing between the initial and final states of the full system, before concluding remarks. In this way, we can elucidate the relations between the energy cost, time constraints and the quantum hypothesis testing, for the information processing in quantum thermodynamic systems. Throughout the paper, we adopt natural units, setting $\hbar = k_B =  1$.

\section{Self-contained quantum universe}
\label{sec:setup}

\begin{figure}[htp!]
    \centering
    \includegraphics[width=.9\columnwidth]{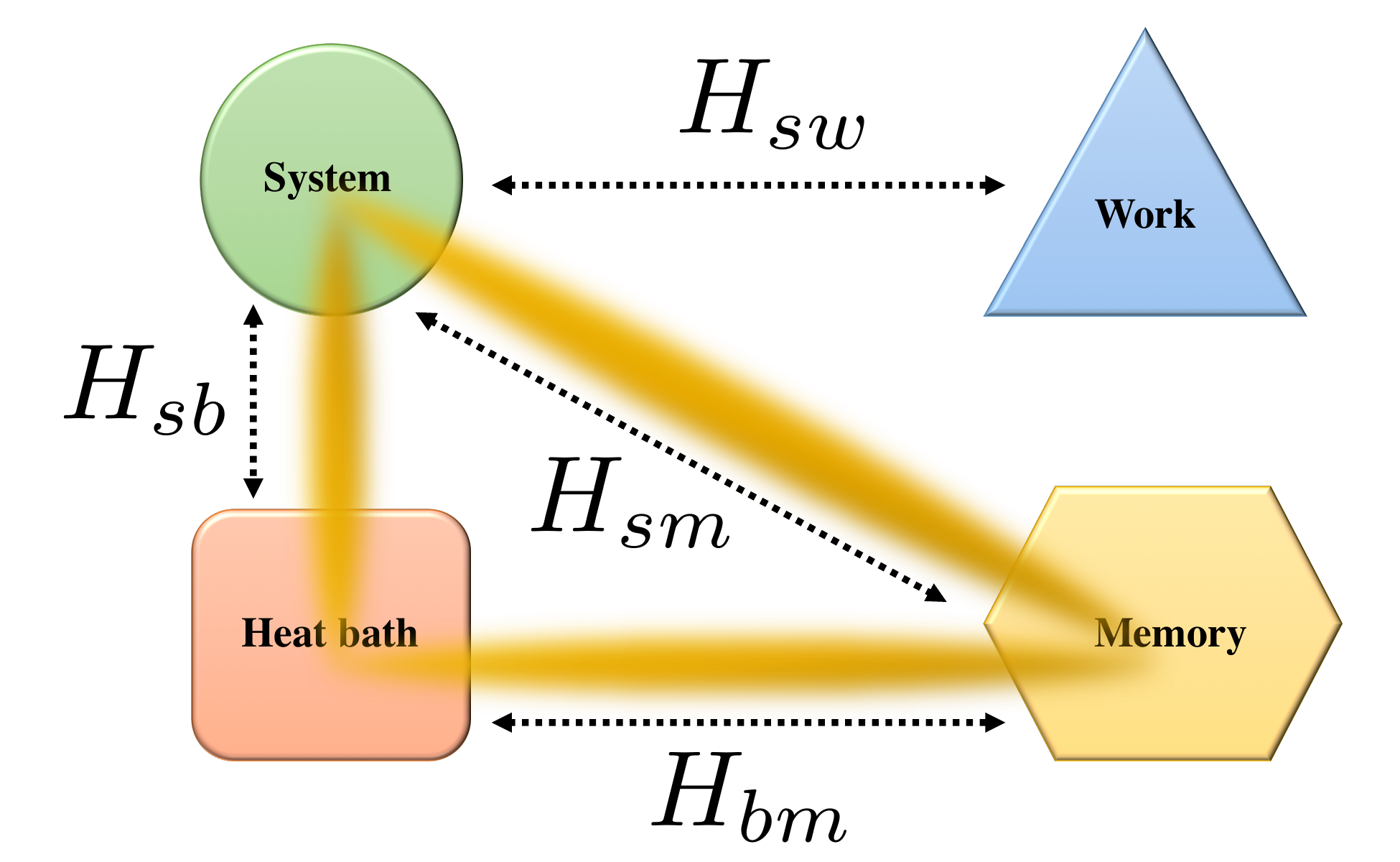}
    \caption{Initially, the heat bath $(\mathcal{H}_b)$, system $(\mathcal{H}_s)$, and memory $(\mathcal{H}_m)$ may share correlations, while the work source $(\mathcal{H}_w)$ is uncorrelated with the remaining subsystems $(\mathcal{H}_{\overline{w}}\equiv\mathcal{H}_b\otimes\mathcal{H}_s\otimes\mathcal{H}_m)$. The system interacts with the bath via $H_{sb}$, with the work source via $H_{sw}$, and with the memory via $H_{sm}$. The bath and memory interact through $H_{bm}$. 
    }
    \label{fig1}
\end{figure}

Following Ref.~\cite{deffner2013information},  {we consider a self-contained ``universe'', which comprises a principal system, a work reservoir (or source), a heat reservoir (or bath), and an information reservoir (or memory), cf. Fig.~\ref{fig1} for a pictorial representation}.
We denote  {by} $\mathcal{H}_b$, $\mathcal{H}_s$, $\mathcal{H}_m$, and $\mathcal{H}_w$ the Hilbert spaces of the heat bath, (principal) system, memory, and work source, respectively, and $d_j\equiv\dim(\mathcal{H}_j)~(j=b,s,m,w)$  {is} their dimension. 

We write $H_s$, $H_b$, $H_m$, and $H_w$ as the bare Hamiltonians of each subsystem. The interactions are described as follows: the system couples to the heat bath via $H_{sb}$, to the memory via $H_{sm}$, and to the work source via $H_{sw}$. The bath also couples to the memory via $H_{bm}$. We restrict our attention to the case where the work source interacts only with the system. Furthermore, we consider an autonomous scenario in which all Hamiltonians are time-independent. The total Hamiltonian is then
\begin{align}
    H_{\tot} = H_0+H_{sb}+H_{sm}+H_{sw}+H_{bm}\,,
\end{align}
where the bare Hamiltonian is defined as
\begin{align}
    H_0 \equiv H_s+H_b+H_m+H_w\,.
\end{align}
 {Accordingly, the ``universe'' evolves under the total unitary operator, $U(t)=e^{-i H_{\tot}t}$.} To satisfy average energy conservation (defined in terms of the bare Hamiltonian $H_0$),  {we further require}
\begin{align}
    [U(t), H_0]=0\,.    \label{eq:EnergyConservation}
\end{align}

We allow the system, bath, and memory to be initially correlated and denote their joint initial state by $\rho_{\overline{w}}$, where $\overline{w}$ denotes the complement of the work source,i.e., $\mathcal{H}_{\overline{w}}\equiv\mathcal{H}_b\otimes\mathcal{H}_s\otimes\mathcal{H}_m$.  The reduced states are given by
\begin{align}
&\text{tr}_{s,m}\left\{\rho_{\overline{w}}\right\}=\rho_b^{\eq}\equiv\frac{1}{Z}e^{-\beta H_b}\label{eq:BathInitial}\\
&\text{tr}_{b,m}\left\{\rho_{\overline{w}}\right\} = \rho_s\label{eq:SystemInitial}\\
&\text{tr}_{s,b}\left\{\rho_{\overline{w}}\right\} = \rho_m\label{eq:MemoryInitial}\,,
\end{align}
where $\beta$ denotes the inverse temperature.  {Note that this temperature can be defined fully within quantum information theory without having to employ classical constructions \cite{Deffner2016NJP}.} 

The initial state of the work source $\rho_w$ is assumed to be uncorrelated with the other subsystems, so the initial state of the total system is $\rho_{\tot}(0)\equiv\rho_{\overline{w}}\otimes\rho_w$. Considering the total evolution from $t=0$ to $t=\tau$, the final state is
\begin{align}
    \rho_{\tot}(\tau) = U(\tau)\left(\rho_{\overline{w}}\otimes\rho_w\right)U\ad(\tau)\,,
\end{align}
and the reduced dynamics for each subsystem is described by
\begin{align}
    \begin{split}
      \mathcal{E}_s(\rho_s) &\equiv \text{tr}_{b,m,w}\left\{U(\tau)\left(\rho_{\overline{w}}\otimes\rho_w\right)U\ad(\tau)\right\}\\
      \mathcal{E}_b(\rho_b^{\eq}) &\equiv \text{tr}_{s,m,w}\left\{U(\tau)\left(\rho_{\overline{w}}\otimes\rho_w\right)U\ad(\tau)\right\}\\
      \mathcal{E}_m(\rho_m) &\equiv \text{tr}_{s,b,w}~\left\{U(\tau)\left(\rho_{\overline{w}}\otimes\rho_w\right)U\ad(\tau)\right\}\\
      \mathcal{E}_w(\rho_w)~ &\equiv \text{tr}_{s,b,m}~\left\{U(\tau)\left(\rho_{\overline{w}}\otimes\rho_w\right)U\ad(\tau)\right\}\,.
    \end{split}
\end{align}

In particular, we impose two conditions on the work source and memory. Following  {and generalizing} Ref.~\cite{deffner2013information}, we want the work source to preserve its von-Neumann entropy under the time evolution $\mathcal{E}_w$ for \textit{any} $\rho_w$,i.e.,
\begin{align}
    \Delta S(\rho_{w})=S(\mathcal{E}_w(\rho_w))-S(\rho_w)=0\,,
\label{eq:WorkEntropy}
\end{align}
where $S(\rho)\equiv-\text{tr}\{\rho\ln\rho\}$ denotes the von Neumann entropy of $\rho$. Then, it is natural for us to consider $\mathcal{E}_{w}$ to be a unitary operation~\cite{he2015entropy}. 

For the memory,  {according to} Ref.~\cite{strasberg2017quantum}, an ideal memory has a Hamiltonian proportional to the identity
\begin{align}
    H_m\propto \id_m\,,
\label{eq:identityHamiltonian}
\end{align}
indicating complete degeneracy. In this case, the memory does not exchange energy with the rest of the system, although its von Neumann entropy may change.

\section{Hamiltonian and time evolution constraints}
\label{sec:unitaryAndhamiltonian}

 {We now have all elements necessary to} derive the requirements placed on the total Hamiltonian such that the dynamics of $\rho_w$ are consistent with that of a work source.
Specifically, we identify the conditions  {for which} Eq.~\eqref{eq:WorkEntropy} is satisfied.

\subsection{General statement}
\label{sec:general_property}
 {We start by inspecting the total time evolution operator $U(t)$.} Using the terminology of quantum resource theory, the work source $\mathcal{H}_w$ must act as a \textit{catalyst} that preserves its initial randomness to adhere to Eq.~\eqref{eq:WorkEntropy}. In Ref.~\cite{Lie21}, Lie and Jeong demonstrated that when $U(\tau)$ is a \textit{catalysis unitary} satisfying 
\begin{align}
\text{tr}_{\overline{w}}\left\{U(\tau)(\rho_{\overline{w}}\otimes\rho_w)U\ad(\tau)\right\} = V_w\rho_w V_w\ad\,,
\label{eq:CatalysisUnitary}
\end{align}
where $V_w$ is a unitary operator on the work source $\mathcal{H}_w$, $U(\tau)$ must be compatible with $\id_{\overline{w}}\otimes\rho_w$, hence
\begin{align}   [U(\tau),\id_{\overline{w}}\otimes\rho_w]=0.
\label{eq:compatible}
\end{align}
Interestingly, its partial transpose $U^{\top_{\overline{w}}}(\tau)$ must be also a unitary operator, where $U^{\top_{\overline{w}}}(\tau)$ denotes the partial transpose of $U(\tau)$ over the subsystem $\mathcal{H}_{\overline{w}}$.
Here, note that the partial transpose $\top_{\overline{w}}$ is taken in the product eigenbasis of the bare Hamiltonians $H_b$, $H_s$ and $H_m$. This choice is physically natural in our framework, since these Hamiltonians define the thermodynamic observables and the tensor decomposition of the total system into the subsystems. As the thermodynamic description is formulated with respect to these bare Hamiltonians, the corresponding energy eigenbasis provides a canonical reference basis for defining the transpose operation.

 {In the classical setting, this effective separation of the dynamics of the work source from the rest of the ``universe'' emerges in the limit in which the inertia of the work source becomes much larger than any other inertia. In the present quantum treatment such a limit is not strictly necessary, since we can leverage the construction of the dynamical maps directly. Moreover, from this construction we obtain a required structure of the Hamiltonian generating the dynamics. Namely, requiring Eq.~\eqref{eq:CatalysisUnitary}, we immediately have that} the dynamics of $\mathcal{H}_{\overline{w}}$  {has to be} unital, as bipartite unitaries whose partial transpose is also unitary are known to induce unital maps independently of the catalyst's state~\cite{Lie21,Deschamps16,Benoist17}. Therefore, writing 
\begin{align}
\mathcal{E}_{\overline{w}}(\rho_{\overline{w}})\equiv
\Tr_w\left\{U(\tau)\left(\rho_{\overline{w}}\otimes\rho_w\right)U\ad(\tau)\right\},
\label{eq:reduced_state_complement}
\end{align}
we have 
\begin{align}
S\left(\mathcal{E}_{\overline{w}}\left(\rho_{\overline{w}}\right)\right)\geq S(\rho_{\overline{w}})\,.
\label{eq:entropy_increase}
\end{align}
Here, note that the local unitary basis transformations on the complement subsystem $\mathcal{H}_{\overline{w}}$ does not change the condition of unitary partial transpose. See Appendix~\ref{app:local_partial_transpose} for details.

Moreover, in our fully autonomous treatment the total Hamiltonian $H_{\tot}$ is \emph{time-independent}, since we take into consideration all of the subsystems involved in the dynamics. Consequently, the unitary operator $U(\tau)$  {can be written as}
\begin{align}
U(\tau)=\sum_{n=0}^{\infty}\frac{(-i \tau)^n}{n!}H_{\tot}^n\,,
\end{align}
 {from which the partial transpose operation is obtained term-by-term,}
\begin{align}
    U^{\top_{\overline{w}}}(\tau)=\sum_{n=0}^{\infty}\frac{(-i \tau)^n}{n!}\Big(H_{\tot}^n\Big)^{\top_{\overline{w}}}.
\label{eq:UtaSeries1}
\end{align}

 Further leveraging that $H_{\tot}^{\top_{\overline{w}}}$ is Hermitian (see Appendix~\ref{app:Hermitian} for details), and $U^{\top_{\overline{w}}}(\tau)$  {is a} unitary operator. Namely,
\begin{align}
    U^{\top_{\overline{w}}}(\tau)=e^{-iH_{\tot}^{\top_{\overline{w}}}\tau}=\sum_{n=0}^{\infty}\frac{(-i\tau)^n}{n!}\left(H_{\tot}^{\top_{\overline{w}}}\right)^n\,.
\label{eq:UtaSeries2}
\end{align}
Comparing Eqs.~\eqref{eq:UtaSeries1} and \eqref{eq:UtaSeries2}, $H_{\tot}$ must satisfy
\begin{align}
\Big(H_{\tot}^n \Big)^{\top_{\overline{w}}}=\left(H_{\tot}^{\top_{\overline{w}}}\right)^n\quad  {(\forall n)}\,.
\label{eq:HnCondition}
\end{align}
In other words, the partial transpose of $H_{\tot}$ must preserve multiplicativity for all powers of $H_{\tot}$. This requirement alone provides a complete list of constraints placed on the total Hamiltonian to enforce that $\rho_w$ evolves in a manner consistent with it acting as a work source.

\subsection{Example}
\label{sec:example}
In the following, based on Eq.~\eqref{eq:HnCondition}, we provide a class of Hamiltonians and corresponding unitaries as an example for $U(\tau)$ to be a catalysis unitary. Let us write
\begin{align}
    H_{\tot} = \sum_{j=1}^{M}A_j\otimes B_j\,,
\end{align}
where $\{A_j\}_{j=1}^{M}$ and $\{B_j\}_{j=1}^{M}$ are nonzero Hermitian operators acting on $\mathcal{H}_{\overline{w}}$ and $\mathcal{H}_{w}$, respectively. 
For the powers of $H_{\tot}$, starting from
\begin{align}
    H_{\tot}^n = \sum_{j_1,\cdots,j_n}(A_{j_1}\cdots A_{j_n})\otimes (B_{j_1}\cdots B_{j_n})\,,
\end{align}
we obtain
\begin{align}
    (H_{\tot}^n)^{\top_{\overline{w}}}=\sum_{j_1,\cdots,j_n}(A_{j_n}^{\top}\cdots A_{j_1}^{\top})\otimes (B_{j_1}\cdots B_{j_n})\,.
\end{align}
The partial transpose of $H_{\tot}$ is given by
\begin{align}
H_{\tot}^{\top_{\overline{w}}} = \sum_{j=1}^{M} A_j^{\top}\otimes B_j\,,
\end{align}
which leads to
\begin{align}
(H_{\tot}^{\top_{\overline{w}}})^n=\sum_{j_1,\cdots,j_n}(A_{j_1}^{\top}\cdots A_{j_n}^{\top})\otimes (B_{j_1}\cdots B_{j_n})\,.
\end{align}
For example, when Eq.~\eqref{eq:HnCondition} holds \textit{any} choice of $\{B_j\}_{j=1}^M$, it is necessary that
\begin{align}
A_{j_1}\cdots A_{j_n}=A_{j_n}\cdots A_{j_1}\,,
\end{align}
for $j_k=1,2,\cdots, M~(\forall k)$. In particular, when $n=2$, this reduces to $[A_{j_1},A_{j_2}]=0~~(\forall j_1,j_2)$. Since all sets $\{A_{j_1}\}_{j_1=1}^{M},\cdots,\{A_{j_n}\}_{j_n=1}^{M}$ are identical to each other, we can simply write
\begin{align}
[A_i,A_j]=0~~(\forall i,j)\,.
\end{align}
This means that all $A_j$ are simultaneously diagonalizable so that they share the same eigenbases
    \begin{align}
        A_j=\sum_{k}
        \lambda_{jk}\dya{k}~~(\forall j)\,.
    \end{align} 
Then, we can write
    \begin{align}   
    U(\tau)
    =\sum_{k}\dya{k}\otimes \Lambda_{k}\,,
    \end{align}
where  
    \begin{align}
    \Lambda_k \equiv \exp\left(-i\tau\sum_{j}\lambda_{jk}B_j\right)
    \end{align}
are unitary operators acting on $\mathcal{H}_w$.
Therefore, in this example, $U(\tau)$ becomes the controlled unitary operation. To ensure the compatibility condition $[U(\tau),\id_{\overline{w}}\otimes\rho_w]=0$, we have 
\begin{align}
[U(\tau),\id_{\overline{w}}\otimes\rho_w]=\sum_{k}\dya{k}\otimes[\Lambda_k,\rho_w]=0\,,
\end{align}
which leads to the condition that $\rho_w$ is the fixed point for all $\Lambda_k$, i.e.,
\begin{align}
\Lambda_k\rho_w\Lambda_k\ad=\rho_w~(\forall k)\,.
\end{align}
In this case, we actually obtain $\mathcal{E}_{\overline{w}}(\rho_{\overline{w}})$ in Eq.~\eqref{eq:reduced_state_complement} satisfying Eq.~\eqref{eq:entropy_increase} and 
\begin{align}
\Tr_{\overline{w}}\left[U(\tau)(\rho_{\overline{w}}\otimes\rho_w)U\ad(\tau)\right]=\rho_w\,, 
\end{align}
where in this case $V_w=\id_w$.
Interestingly, we want to note that the corresponding
$U(\tau)$ has the properties as the unitary for which the converse statement of the catalytic entropy conjecture holds true~\cite{boes2019von, wilming2021entropy, wilming2022correlations}.

\section{Laws of Thermodynamics}

 {Now that we have constructed the Hamiltonian from its corresponding dynamics, we formulate the statements of the laws of thermodynamics.}

\label{sec:laws}
\subsection{First law of thermodynamics}

To state the first law of thermodynamics, we must have expressions for heat, work, and internal energy.  {As always, we} take heat $Q$ to be the energy dissipated from the heat bath, 
\begin{align}
Q\equiv\text{tr}\left\{\left(\rho_b^{\eq}-\mathcal{E}_b\left(\rho_b^{\eq}\right)\right)H_b\right\}\,,
\label{eq:Heat}
\end{align}
work $W$ to be the energy dissipated from the work source, 
\begin{align}
    W\equiv\text{tr}\left\{\left(\rho_w-\mathcal{E}_w\left(\rho_w\right)\right)H_w\right\}\,.
\label{eq:Work}
\end{align}
 {The} internal energy change $\Delta E$ as the system’s energy change during evolution  {simply is}, 
\begin{align}
\Delta E\equiv \text{tr}\left\{\left(\mathcal{E}_s\left(\rho_s\right)-\rho_s\right)H_s\right\}\,.
\label{eq:InterEnergy}
\end{align}
Following Refs.~\cite{deffner2013information,strasberg2017quantum}, and using Eq.~\eqref{eq:identityHamiltonian}, we  immediately have for the memory~\footnote{If the memory Hamiltonian is nondegenerate, energy exchange between the system and memory must be taken into account. As a result, the simplest forms of the first- and second-law bounds, including Landauer principle, require modification by incorporating the memory’s energy change into the effective heat. While this leads to quantitative corrections, namely making the minimal work for information erasure depend on both entropy and energy changes, the overall structure of the bounds and their core conceptual insights remain unchanged.}
\begin{align}
    0 = \text{tr}\left\{\left(\rho_m-\mathcal{E}_m\left(\rho_m\right)\right)H_m\right\}\,.
\label{eq:Memory}
\end{align}
Combining Eqs.~\eqref{eq:EnergyConservation}, \eqref{eq:Heat}, \eqref{eq:Work}, \eqref{eq:InterEnergy}, and \eqref{eq:Memory}, the first law of thermodynamics  {then naturally emerges},
\begin{align}
    \Delta E = Q + W\,.
    \label{eq:1stLaw}
\end{align}
 {We emphasize that there is no energetic contribution from the memory, since its reduced Hamiltonian is fully degenerate and, hence, the memory cannot exchange energy with the other subsystems directly. However, as we will see in the formulation of the second law, its change in entropy does lead to an exchange of energy among the other subsystems.}

\subsection{Second law of thermodynamics}

 {To formulate the second law of thermodynamics, we note that the heat bath can be assumed to be in a Gibbs state \eqref{eq:BathInitial}, and we have}
\begin{align}
    \ln\rho_b^{\eq}=-\beta H_b-\ln Z_b\,.
\end{align}
Using Eq.~\eqref{eq:Heat}, this gives
\begin{align}
    \beta Q 
    = S(\rho_b^{\eq})+\text{tr}\left\{\mathcal{E}_b\left(\rho_b^{\eq}\right)\ln \rho_b^{\eq}\right\}\,.
\end{align} 
We now define 
\begin{align}
\rho_{\tot}(\tau) &\equiv U(\tau)\left(\rho_{\overline{w}}\otimes\rho_{w}\right)U\ad(\tau),
\label{eq:finalstate}\\
\sigma_{\tot}(\tau) &=\rho_b^{\eq}\otimes \mathcal{E}_s(\rho_s)\otimes\mathcal{E}_m(\rho_m)\otimes\mathcal{E}_{w}(\rho_w)\,.
\label{eq:productstate}
\end{align}
Further, we write $S(\rho\,||\,\sigma)\equiv-S(\rho)-\text{tr}\{\rho\ln\sigma\}$ for the quantum relative entropy of $\rho$ with respect to $\sigma$. Then, defining 
\begin{align}
    \Delta S(\rho_s)&\equiv S(\mathcal{E}_s(\rho_s))-S(\rho_s)\\
    \Delta S(\rho_m)&\equiv S(\mathcal{E}_m(\rho_m))-S(\rho_m)\,,
\end{align}
and the quantum relative entropy difference,
\begin{align}
    \!\!\!\!\!\!\delta_{\text{rel}}(\tau)\equiv S(\rho_{\tot}(\tau)\,||\,\sigma_{\tot}(\tau))-S(\rho_{\tot}(0)\,||\,\sigma_{\tot}(0))\,,
\end{align}
we  {obtain} the second law of thermodynamics as
\begin{align}
    \Delta S(\rho_s)+\Delta S(\rho_m)-\beta Q = \delta_{\text{rel}}(\tau)\,.
\label{eq:2ndLaw}
\end{align}
See Appendix~\ref{app:2ndLawProof} for a more detailed derivation. 

 {Observe that Eq.~\eqref{eq:2ndLaw} asserts that the irreversible entropy production, that is the difference between change in entropy and heat exchanged with the heat bath, is governed by the change of entropy in the memory. Thus, Eq.~\eqref{eq:2ndLaw} can be read a as quantum generalization of Landauer's principle for autonomous quantum information processing. This can be made even more precise by defining} the effective heat as
\begin{align}
    Q_{\text{eff}}=Q-\beta^{-1}S(\rho_{\tot}(0)\,||\,\sigma_{\tot}(0))\,,
\end{align}
 {from which we obtaion}
\begin{align}
    \Delta S(\rho_s)+\Delta S(\rho_m)-\beta Q_{\text{eff}}=S(\rho_{\tot}(\tau)\,||\,\sigma_{\tot}(\tau))\,.
\label{eq:Landuer_equality}
\end{align}
Due to the non-negativity of the quantum relative entropy, $S(\rho_{\tot}(\tau)\,||\,\sigma_{\tot}(\tau))\geq 0$, we  {then have the quantum} Landauer bound,
\begin{align}
    \Delta S(\rho_s)+\Delta S(\rho_m)\geq \beta Q_{\text{eff}}\,.
    \label{eq:Landauer_bound}
\end{align}
Here, note that $S(\rho_{\tot}(0)\,||\,\sigma_{\tot}(0))=S(\rho_b^{\eq})+S(\rho_s)+S(\rho_m)-S(\rho_{\overline{w}})$ is equivalent to the tripartite quantum mutual information contained in the initial state  $\rho_{\overline{w}}$~\cite{Watanabe1960}, meaning that the Landauer bound depends on the initial correlations.

The correlation-dependent terms in the derived Landauer relations have a clear physical meaning as follows. As an example, in the picture of Maxwell’s demon~\cite{maruyama2009maxwell,bennett2003notes}, these terms quantify how pre-existing correlations affect the energetic cost of information processing. The correlations between the memory and system can either enhance efficiency, if they provide useful prior knowledge, or increase the cost, if they introduce misaligned uncertainty. Likewise, correlations with the bath encode hidden information about energy exchanges. Overall, these terms show that initial correlations act as a key resource or constraint, modifying thermodynamic bounds while preserving their fundamental structure.

\section{Quantum Thermodynamic Speed Limit}
\label{sec:QTSL}

 {The autonomous Hamiltonian approach to quantum information processing has the added benefit that we also have full access to the dynamics. In particular, we can now also formulate statements for the rate of entropy change, the entropy production, which is upper bound by the QSL \cite{Deffner2010PRL}.}

 {While there are many formulations for open quantum systems, the version of the QSL from Ref.~\cite{deffner2020quantum} is particularly convenient for our present purposes. This version of the QSL is expressed in terms of the Schatten $p$\,-norm, which} enables flexible analyses of numerical stability and spectral sensitivity. For instance, when $p=2$, the corresponding Hilbert--Schmidt distance can be efficiently obtained via the SWAP test by computing the state overlap~\cite{cincio2018learning}; therefore, the quantity based on the Schatten $p$-norm can be both numerically tractable and experimentally implementable.

Given a operator $A$ acting on a $d$-dimensional Hilbert space, its Schatten $p$\,-norm 
\begin{align}
\norm{A}_{p}\equiv \left(\text{tr}\left\{\left(A\ad A\right)^{\frac{p}{2}}\right\}\right)^{\frac{1}{p}}
\end{align}
satisfies the following inequality~\cite{horn2012matrix},
\begin{align}
    \norm{A}_{p}\leq \norm{A}_{q}\leq  d^{\frac{1}{q}-\frac{1}{p}}\norm{A}_{p}~~(0<q\leq p)\,.
\label{eq:p-norm-inequality}
\end{align}

Following Refs.~\cite{deffner2020quantum,deffnerquantum2013, deffner2017geometric}, we define the time-averaged Schatten $p$\,-norm as 
\begin{align}
    \Lambda_{s,m}^{(p)}(\tau)\equiv\frac{1}{\tau}\int_{0}^{\tau}\norm{\partial_{t}\rho_{s,m}(t)}_{p}dt\,,
    \label{eq:time-averaged-p-norm}
\end{align}
where $\partial_t\equiv\partial/\partial t$, and the QSL time as 
\begin{align}
T_{s,m}^{(p)}\equiv\frac{\ell_p(\mathcal{E}_{s,m}(\rho_{s,m}), \rho_{s,m})}{\Lambda_{s,m}^{(p)}(\tau)}\,,
    \label{eq:qsl-time}
\end{align}
where 
\begin{align}
    \ell_p(\rho_1,\rho_2)\equiv\norm{\rho_1-\rho_2}_p
    \label{eq:SchattenDistance}
\end{align}
denotes the Schatten $p$\,-distance between two quantum states $\rho_1$ and $\rho_2$~\cite{dambal2025harnessing}. 

From Fannes inequality~\cite{fannes1973continuity,audenaert2007sharp,Nielsen,deffner2020quantum} and Eq.~\eqref{eq:p-norm-inequality}, we  {then} have 
\begin{align}
\begin{split}
    \!\!\!\!\abs{\Delta S_s  +\Delta S_m }&\leq d_s  ^{1-\frac{1}{p}}\ln(d_s)\ell_p(\mathcal{E}_s(\rho_s),\rho_s)\\
    &+d_m^{1-\frac{1}{p}}\ln(d_m)\ell_p(\mathcal{E}_m(\rho_m),\rho_m)+\frac{2}{e}\,.
\end{split}
\end{align}
By using the dimension relation 
\begin{align}
    d\equiv\dim(\mathcal{H}_s\otimes\mathcal{H}_m)=d_sd_m\,,
\label{eq:dimensions}
\end{align}
we define 
\begin{align}
    \Lambda_{p}^{\star}\equiv \frac{d_s^{1-\frac{1}{p}}\ln(d_s)\Lambda_s^{(p)}+d_m^{1-\frac{1}{p}}\ln(d_m)\Lambda_m ^{(p)}}{d^{1-\frac{1}{p}}\ln(d)},
\end{align}
and the effective QSL time
\begin{align}
T^{\star}_{p}\equiv\frac{d_s^{1-\frac{1}{p}}\ln(d_s)\Lambda_s^{(p)}T_s^{(p)}+d_m ^{1-\frac{1}{p}}\ln(d_m)\Lambda_m^{(p)}T_m^{(p)}}{d_s^{1-\frac{1}{p}}\ln(d_s)\Lambda_s^{(p)}+d_m^{1-\frac{1}{p}}\ln(d_m)\Lambda_m ^{(p)}}\,.
\label{eq:QTSL}
\end{align}
Here, we refer to $T^{\star}_{p}$ as the \textit{Quantum Thermodynamic Speed Limit} (QTSL) time (of order $p$); cf. Appendix~\ref{app:example} for two examples illustrating the quantity $T_1^{\star}$. 

 {This thermodynamic version of the quantum speed limit determines the average rate with which thermodynamic resources can be exchanged between the different subsystems. In particular, it bounds the rate with which information \emph{about} the principal system can be encoded in the memory. This interpretation becomes more stringent by reinspecting the quantum Landauer bound \eqref{eq:Landauer_bound}. In particular, we can} write 
\begin{align}
    \abs{\Delta S_s  +\Delta S_m }\leq d^{1-\frac{1}{p}}\ln(d)\Lambda^{\star}_pT^{\star}_{p}+\frac{2}{e}\,.
\end{align}
Following Ref.~\cite{deffner2020quantum}, using the quantum information theoretic version of Bremermann-Bekenstein bound of order $p$ defined as 
\begin{align}
    B^{\star}_{p}(\tau)\equiv\ln(d)d^{1-\frac{1}{p}}\Lambda^{\star}_{p}(\tau)\,,
\end{align}
from Eq.~\eqref{eq:Landuer_equality}, we obtain
\begin{align}
    S(\rho_{\tot}(\tau)\,||\,\sigma_{\tot}(\tau))+\beta Q_{\text{eff}} \leq T^{\star}_{p}B^{\star}_{p}(\tau)+\frac{2}{e}\,.
\label{eq:LandauerBoundAndQTSL_p}
\end{align}
 {Equation~\eqref{eq:LandauerBoundAndQTSL_p} can be interpreted as a dynamical version of the Landauer bound for the communication of information between a thermodynamic system and a quantum memory.} Importantly, the two bounds  {Eq.~\eqref{eq:Landauer_bound} and Eq.~\eqref{eq:LandauerBoundAndQTSL_p} are complementary: the Landauer bound \eqref{eq:Landauer_bound} provides a lower bound on the change in entropy of the system and memory, while the QTSL bound Eq.~\eqref{eq:LandauerBoundAndQTSL_p}} provides an upper bound.

\subsection{ {Quantum hypothesis testing}}

 {Finally we} discuss the operational meaning of $T^{\star}_p$ in the context of quantum hypothesis testing, as described by the quantum Stein lemma~\cite{Ogawa00, Brandao10}.  
Let $p_n(\tau)$ be the minimum type-I\!I error probability of a testing protocol,i.e., the probability that the positive operator-valued measure (POVM) measurement indicates $\rho_{\tot}(\tau)$ when the true state is $\sigma_{\tot}(\tau)$, given $n$ copies to test. Then
\begin{align}
S(\rho_{\tot}(\tau)\,||\,\sigma_{\tot}(\tau)) = -\lim_{n\to\infty}\frac{1}{n}\ln p_n(\tau)\,. 
\end{align}
From Eq.~\eqref{eq:LandauerBoundAndQTSL_p}, we obtain
\begin{align}
    -\lim_{n\to\infty}\frac{1}{n}\ln p_n(\tau)\leq T^{\star}_pB^{\star}_p(\tau)-\beta Q_{\text{eff}}+\frac{2}{e}\,.
\label{eq:QTSLandQuantumHypothesisTesting}
\end{align}
Thus, QTSL time provides an upper bound on the scaling behavior in quantum hypothesis testing.

\section{ {Concluding remarks}}
\label{sec:conc}

In conclusion, we have extended the classical framework of information processing in thermodynamic systems from Ref.~\cite{deffner2013information} to the quantum domain. Focusing on a composite system comprising a principal system, heat bath, memory, and work source within an autonomous Hamiltonian setting, we derived constraints on the total Hamiltonian required for  thermodynamically consistent description of quantum information processing. 
By generalizing to scenarios where the system, bath, and memory may initially be correlated, we extended the second law of thermodynamics to account for information-processing effects in autonomous quantum systems. We proposed that the total unitary evolution should admit a unitary partial transpose and be compatible with the state of the work source, ensuring that the work source acts catalytically and preserves its von Neumann entropy. This condition enforces commutativity among operators acting on the joint system and establishes the Hamiltonian structure. Moreover, by analyzing the quantum speed limit of the system and memory, we introduced the concept of a quantum thermodynamic speed limit characterized by the Schatten norm and demonstrated its connection to the quantum Landauer bound. Finally, considering quantum Stein’s lemma, we provided an operational interpretation of this speed limit in the context of quantum hypothesis testing. 
Together, these results establish a unified framework for understanding the interplay between energetic and temporal constraints in quantum information processing within general quantum thermodynamic systems.

\acknowledgements{We thank Seok Hyung Lie for a helpful discussion. S-I.T, E.D. and A.S acknowledges U.S. NSF under Grant No. OSI-2328774. A.S. also acknowledges PHY-2425180 and Cooperative Agreement PHY-2019786.  A.T. is supported by the U.S DOE under the LDRD program at Los Alamos. S.D. acknowledges support from the John Templeton Foundation under Grant No. 63626. This work was supported by the U.S. Department of Energy, Office of Basic Energy Sciences, Quantum Information Science program in Chemical Sciences, Geosciences, and Biosciences, under Award No. DE-SC0025997.}

\onecolumngrid

\appendix

\section{Invariance of Unitary Partial Transpose under Local Basis Transformation}
\label{app:local_partial_transpose}

We demonstrate that the unitary partial transpose condition is stable under local unitary basis transformations on the complement subsystem $\mathcal{H}_{\overline{w}}$. Specifically, if the basis of $\mathcal{H}_{\overline{w}}$ is changed by a unitary $\Gamma_{\overline{w}}$, the global unitary transforms as 
    \begin{align}
        U'=(\Gamma_{\overline{w}}\otimes \id_w)U(\Gamma_{\overline{w}}\ad\otimes\id_w)\,.
    \end{align}
    Now, writing 
    \begin{align}
        U = \sum_{ijk\ell}u_{ijk\ell}\ketbra{a_i}{a_j}_{\overline{w}}\otimes\ketbra{b_k}{b_{\ell}}_{w}\,,
    \end{align}
    where $\ket{a_i}$ and $\ket{b_k}$ denote the bases of $\mathcal{H}_{\overline{w}}$ and $\mathcal{H}_w$, 
    then, we have 
    \begin{align}
        U'=\sum_{ijk\ell}u_{ijk\ell}\Gamma_{\overline{w}}\ketbra{a_i}{a_j}\Gamma_{\overline{w}}\ad\otimes \ketbra{b_k}{b_{\ell}}\,.
    \end{align}
    Since 
    \begin{align}
        U^{\top_{\overline{w}}} = \sum_{ijk\ell}u_{ijk\ell}\ketbra{a_j}{a_i}_{\overline{w}}\otimes\ketbra{b_{k}}{b_{\ell}}_{w}
    \end{align}
    and 
    \begin{align}
(U')^{\top_{\overline{w}}}
        = \sum_{ijk\ell}(\Gamma_{\overline{w}}^{\dagger})^{\top}\ketbra{a_j}{a_i}\Gamma_{\overline{w}}^{\top}\otimes\ketbra{b_k}{b_{\ell}}=\left(\Gamma_{\overline{w}}^{*}\otimes\id_{w}\right)\left(\sum_{ijk\ell}u_{ijk\ell}\ketbra{a_j}{a_i}_{\overline{w}}\otimes\ketbra{b_k}{b_{\ell}}_w\right)(\Gamma_{\overline{w}}^{\top}\otimes\id_{w})\,,
    \end{align}
    we have 
    \begin{align}
        (U')^{\top_{\overline{w}}}=(\Gamma_{\overline{w}}^*\otimes\id_w)U^{\top_{\overline{w}}}(\Gamma_{\overline{w}}^{\top}\otimes\id_w)\,.
    \end{align}
    Since left and right multiplication by unitaries (i.e., $V_{\overline{w}}^*$ and $V_{\overline{w}}^{\top}$) preserves unitary, the unitarity of $U^{\top_{\overline{w}}}$ is invariant under such local basis changes. 

\section{Hermiticity of Partially Transposed Hamiltonians}
\label{app:Hermitian}
Let $H$ be a Hamiltonian acting on a composite Hilbert space $\mathcal{H}_{\overline{w}}\otimes\mathcal{H}_{w}$. In the computational basis, we can express
\begin{align} H=\sum_{ijk\ell}h_{ijk\ell}\ketbra{i}{j}\otimes\ketbra{k}{\ell}\,.
\end{align}
Then, we can write 
\begin{align}
H\ad=\sum_{ijk\ell}h_{ijk\ell}^*\ketbra{j}{i}\otimes\ketbra{\ell}{k}
=\sum_{ijk\ell}h_{ji\ell k}^*\ketbra{i}{j}\otimes\ketbra{k}{\ell}\,,
\end{align}
where $*$ represents the complex conjugate. Since $H$ is Hermitian, i.e., $H=H\ad$, it follows that 
\begin{align}
h_{ijk\ell}=h_{ji\ell k}^*\,.
\label{app:eq:elementsHamiltonian}
\end{align}

Applying the partial transpose $\top_{\overline{w}}$ over subsystem $\mathcal{H}_{\overline{w}}$, we obtain
\begin{align}
H^{\top_{\overline{w}}} = \sum_{ijk\ell}h_{ijk\ell}\ketbra{j}{i}\otimes\ketbra{k}{\ell}\,.
\end{align}
Its conjugate transpose is
\begin{align}
\left(H^{\top_{\overline{w}}}\right)\ad 
=\sum_{ijk\ell}h_{ijk\ell}^*\ketbra{i}{j}\otimes\ketbra{\ell}{k}
=\sum_{ijk\ell}h_{ji\ell k}^*\ketbra{j}{i}\otimes\ketbra{k}{\ell}\,.
\end{align}
From Eq.~\eqref{app:eq:elementsHamiltonian}, we can finally find that the partial transpose preserves Hermiticity
\begin{align}
H^{\top_{\overline{w}}}=\left(H^{\top_{\overline{w}}}\right)\ad\,.
\end{align}

\section{Derivation of Eq.~\eqref{eq:2ndLaw}}
\label{app:2ndLawProof}
From Eqs.~\eqref{eq:finalstate}, \eqref{eq:productstate} and ~\eqref{eq:WorkEntropy}, we derive Eq.~\eqref{eq:2ndLaw} as follows, 
\begin{align}
    \begin{split}
        S&(\rho_{\tot}(\tau)\,||\,\sigma_{\tot}(\tau))-
        S(\rho_{\tot}(0)\,||\,\sigma_{\tot}(0))\\
        =&-S(\rho_{\tot}(\tau))-\text{tr}\left\{\rho_{\tot}(\tau)\ln\sigma_{\tot}(\tau)\right\}
        +S(\rho_{\tot}(0))+\text{tr}\left\{\rho_{\tot}(0)\ln\sigma_{\tot}(0)\right\}\\
        =&-\text{tr}\left\{U(\tau)(\rho_{\overline{w}}\otimes\rho_w)U\ad(\tau)\ln\left(\rho_b^{\eq}\otimes\mathcal{E}_s(\rho_s)\otimes\mathcal{E}_m(\rho_m)\otimes\mathcal{E}_w(\rho_w)\right)\right\}
        +\text{tr}\left\{(\rho_{\overline{w}}\otimes\rho_w)\ln\left(\rho_b^{\eq}\otimes\rho_s\otimes\rho_m\otimes\rho_w\right)\right\}\\
        =&-\text{tr}\{\mathcal{E}_b(\rho_b^{\eq})\ln\rho_b^{\eq}\}+S(\mathcal{E}_s(\rho_s))
        +S(\mathcal{E}_m(\rho_m))-S(\rho_b^{\eq})-S(\rho_b)-S(\rho_m)\\
        =&\Delta S(\rho_s)+\Delta S(\rho_m)-\left(\text{tr}\{\mathcal{E}_b(\rho_b^{\eq})\ln\rho_b^{\eq}\}+S(\rho_b^{\eq})\right)\\
        =&\Delta S(\rho_s)+\Delta S(\rho_m)-\beta Q\,,
    \end{split}
\end{align}
where we used  $S(\rho_{\tot}(\tau))=S(\rho_{\tot}(0))$ due to the unitary invariance of the von-Neumann entropy.

\section{Examples} \label{app:example}

We analyze the behavior of the QTSL of order~$1$ for several representative examples. 
For $p=1$, the QTSL time is expressed as
\begin{align} 
\!\!\!\!T^{\star}_1 = \frac{\ln(d_s)\ell_1(\mathcal{E}_s(\rho_s),\rho_s)+\ln(d_m)\ell_1(\mathcal{E}_m(\rho_m),\rho_m)}{\ln(d_s)\Lambda^{(1)}_s+\ln(d_m)\Lambda^{(1)}_m}\,,
\label{eq:T1}
\end{align}
where
\begin{align} 
\Lambda^{(1)}_{s,m}(\tau)=\frac{1}{\tau}\int_{0}^{\tau}\norm{\partial_t\rho_{s,m}(t)}_1dt\,.
\end{align}

We consider a four-qubit composite system $\mathcal{H}_b\otimes\mathcal{H}_s\otimes\mathcal{H}_m\otimes\mathcal{H}_w$ with dimensions $d_b=d_s=d_m=d_w=2$. 
The Pauli matrices are given by
\begin{align} 
X=
\begin{pmatrix} 
0 & 1\\ 
1 & 0 
\end{pmatrix}\,,~
Y=
\begin{pmatrix} 
0 & -i\\ 
i & 0 
\end{pmatrix}\,,~
Z=
\begin{pmatrix} 
1 & 0\\ 
0 & -1 
\end{pmatrix}\,,
\end{align}
and the computational basis states are
\begin{align} 
\ket{0}=\begin{pmatrix}1\\0\end{pmatrix}\,,~
\ket{1}=\begin{pmatrix}0\\1\end{pmatrix}\,.
\end{align}
We further define the states
\begin{align} 
\ket{+}\equiv\frac{\ket{0}+\ket{1}}{\sqrt{2}}\,,~
\ket{-}\equiv\frac{\ket{0}-\ket{1}}{\sqrt{2}}\,.
\end{align}

The bare Hamiltonians are specified as
\begin{align} 
\begin{split} 
H_s&=\id_b\otimes Z_s\otimes\id_m\otimes\id_w\,,\\
H_b&=Z_b\otimes\id_s\otimes\id_m\otimes\id_w\,,\\
H_m&=\id_b\otimes\id_s\otimes\id_m\otimes\id_w\,,\\
H_w&=\id_b\otimes\id_s\otimes\id_m\otimes Z_w\,,
\end{split} 
\label{eq:BareHamiltonianExample}
\end{align}
and the interaction Hamiltonians are
\begin{align} 
\begin{split} 
H_{sw}&=\id_b\otimes Z_s\otimes\id_m\otimes Z_w\,,\\
H_{sm}&=\id_b\otimes Z_s\otimes Z_m\otimes\id_w\,,\\
H_{sb}&=Z_b\otimes Z_s\otimes\id_m\otimes\id_w\,,\\
H_{bm}&=Z_b\otimes\id_s\otimes Z_m\otimes\id_w\,.
\end{split} 
\label{eq:InteractionHamiltonianExample}
\end{align}
The initial state of the total system is taken as a product state
\begin{align} 
\rho_{\tot}(0) = \rho_b\otimes\rho_{sm}\otimes\rho_w\,,
\end{align}
where the work and heat bath are initially prepared in  
\begin{align}
\begin{split}
\rho_w&=\dya{1}\\
\rho_{b}&=\frac{e^{-Z}}{\text{tr}\{e^{-Z}\}}\,.
\end{split}
\label{eq:InitialStateWBExample}
\end{align}

\subsection{C-maybe state} 
\label{app:CMAYBEExplicit}
We consider the case where the system and memory are initially prepared in $\mathcal{H}_s\otimes\mathcal{H}_m$ is
\begin{align}
\rho_{sm}=\dya{\theta}\,,
\end{align}
where we define
\begin{align}
\ket{\theta}&\equiv\frac{\ket{-}_s\otimes\ket{-}_m+\ket{+}_s\otimes(\cos(\theta)\ket{+}_m+\sin(\theta)\ket{-}_m)}{\sqrt{2}}\,.
\label{eq:cmaybe}
\end{align}
We refer to this as the \textit{C-maybe state}, as it can be generated by the so-called C-maybe gate~\cite{touil2022eavesdropping,girolami2022redundantly} acting on $\mathcal{H}_s\otimes\mathcal{H}_m$ in the $\{\ket{+},\ket{-}\}$ basis. 
This gate plays a significant role in the framework of quantum Darwinism~\cite{touil2022eavesdropping,girolami2022redundantly,Zurek_2025,zurek2009quantum,zurek2022quantum,duruisseau2023pointer,Touil2024QuantumBranching,doucet2024classifying,doucet2025compatibility,Touil2025ConsensusPreprint}.

For the C-maybe state defined in Eq.~\eqref{eq:cmaybe}, under the dynamics generated by the Hamiltonians in Eqs.~\eqref{eq:BareHamiltonianExample} and~\eqref{eq:InteractionHamiltonianExample}, we obtain
\begin{align}
\rho_s&= 
\begin{pmatrix} 
\cos^2\!\left(\frac{\theta}{2}\right)&0\\ 
0&\sin^2\!\left(\frac{\theta}{2}\right)
\end{pmatrix},\\
\rho_m&=\frac{1}{2}
\begin{pmatrix}
1+\sin(\theta)\cos(\theta)&\cos^2(\theta)\\
\cos^2(\theta)&1-\sin(\theta)\cos(\theta)
\end{pmatrix},\\
\mathcal{E}_s(\rho_s)&=\frac{1}{4}
\begin{pmatrix}
4\cos^2\!\left(\frac{\theta}{2}\right)&i\sin(2\theta)\sin(2\tau)\\
-i\sin(2\theta)\sin(2\tau)&4\sin^2\!\left(\frac{\theta}{2}\right)
\end{pmatrix},\\
\mathcal{E}_m(\rho_m)&=\frac{1}{2}
\begin{pmatrix}
1+\sin(\theta)\cos(\theta)&\frac{e^{-4i\tau}(1+e^{2+4i\tau})\cos(\theta)\left(1+e^{4i\tau}(-1+\cos(\theta))+\cos(\theta)\right)}{2(1+e^2)}\\
\frac{e^{-4i\tau}(e^2+e^{4i\tau})\cos(\theta)\left(-1+\cos(\theta)+e^{4i\tau}(1+\cos(\theta))\right)}{2(1+e^2)}&1-\sin(\theta)\cos(\theta)
\end{pmatrix}.
\end{align}

The corresponding trace norms are
\begin{align}
\begin{split}
\ell_1(\mathcal{E}_s(\rho_s),\rho_s)&=\frac{1}{2}\abs{\sin(2\theta)\sin(2\tau)}
=\abs{\sin(\theta)\cos(\theta)\sin(2\tau)}\,,\\
\ell_1(\mathcal{E}_m(\rho_m),\rho_m)&=\frac{\abs{\cos(\theta)\sin(2\tau)}}{1+e^2}
\sqrt{2(1-e^4)\cos(\theta)+\frac{1+e^4}{2}(3+2\cos(2\theta))+2e^2\sin^2(\theta)\cos(4\tau)}\,.
\end{split}
\end{align}

The time-averaged trace norms take the following forms.
For $\Lambda_s^{(1)}(\tau)$, we find
\begin{align}
\Lambda_s^{(1)}(\tau)=\frac{\abs{\sin(2\theta)}}{\tau}\int_{0}^{\tau}\abs{\cos(2t)}dt\,.
\end{align}
By introducing $x=t+\frac{\pi}{4}$, we can rewrite
\begin{align}
\int_{0}^{\tau}\abs{\cos(2t)}dt
=\int_{\frac{\pi}{4}}^{\tau+\frac{\pi}{4}}\abs{\sin(2x)}dx\,.
\label{eq:integralcos2t_1}
\end{align}
The integral over one period $x\in\left[\frac{(N-1)\pi}{2},\frac{N\pi}{2}\right]~(N=1,2,3,\cdots)$ satisfies
\begin{align}
\int_{\frac{(N-1)\pi}{2}}^{\frac{N\pi}{2}}\abs{\sin(2x)}dx
=\int_{0}^{\frac{\pi}{2}}\sin(2x)dx=1\,.
\end{align}
When $\frac{N\pi}{2}\leq \tau+\frac{\pi}{4}<\frac{(N+1)\pi}{2}$, we have $N = \left\lfloor\frac{2\tau}{\pi}+\frac{1}{2}\right\rfloor$, 
where $\lfloor r\rfloor$ denotes the integer part of a real number $r$.
Using the symmetry of $\abs{\sin(2x)}$, we obtain
\begin{align}
\begin{split}
\int_{\frac{\pi}{4}}^{\tau+\frac{\pi}{4}}\abs{\sin(2x)}\,dx=N-\frac{1}{2}+\int_{0}^{\tau+\frac{\pi}{4}-\frac{N\pi}{2}}\sin(2x)\,dx=N+\frac{1}{2}(-1)^N\sin(2\tau)\,.
\end{split}
\label{eq:integralcos2t_2}
\end{align}
Therefore,
\begin{align}
\Lambda_s  ^{(1)}(\tau)
=\frac{\abs{\sin(\theta)\cos(\theta)}}{\tau}\left(2N+(-1)^N\sin(2\tau)\right)\,.
\end{align}

For $\Lambda_m ^{(1)}(\tau)$, we have
\begin{align}
\Lambda_m ^{(1)}(\tau)
=\frac{2\abs{\cos(\theta)}}{(1+e^2)\tau}
\int_{0}^{\tau}\sqrt{(1+\cos(\theta))^2+e^{4}(1-\cos(\theta))^2+2e^2\sin^2(\theta)\cos(8t)}\,dt\,.
\end{align}
When $a+b\neq 0$, the following relation holds:
\begin{align}
\int_0^{\tau}\sqrt{a+b\cos(2mt)}dt
=\frac{\sqrt{a+b}}{m}\int_{0}^{m\tau}\sqrt{1-\frac{2b}{a+b}\sin^2(x)}\,dx=\frac{\sqrt{a+b}}{m}\,\mathfrak{E}\!\left[m\tau\,\Bigg|\,\frac{2b}{a+b}\right],
\label{eq:EllipticFunction}
\end{align}
where $\mathfrak{E}$ denotes the incomplete elliptic integral of the second kind~\cite{byrd2013handbook}, defined as
\begin{align}
\mathfrak{E}\left[\phi\,|\,k^2\right]\equiv\int_{0}^{\phi}\sqrt{1-k^2\sin^2(x)}\,dx\,.
\end{align}
Hence, we obtain
\begin{align}
\begin{split}
\int_{0}^{\tau}&\sqrt{(1+\cos(\theta))^2+e^{4}(1-\cos(\theta))^2+2e^2\sin^2(\theta)\cos(8t)}\,dt\\
&=\frac{1+e^2+(1-e^2)\cos(\theta)}{4}\,
\mathfrak{E}\!\left[4\tau\,\Bigg|\,\left(\frac{2e\sin(\theta)}{1+e^2+(1-e^2)\cos(\theta)}\right)^2\,\right]\,.
\end{split}
\end{align}
Therefore,
\begin{align}
\Lambda_m^{(1)}(\tau)
=\frac{\abs{\cos(\theta)}\left(1+e^2+(1-e^2)\cos(\theta)\right)}{2(1+e^2)\tau}\,
\mathfrak{E}\!\left[4\tau\,\Bigg|\,\left(\frac{2e\sin(\theta)}{1+e^2+(1-e^2)\cos(\theta)}\right)^2\,\right].
\end{align}

\begin{figure}[htp!]
\centering
\includegraphics[width=0.6\columnwidth]{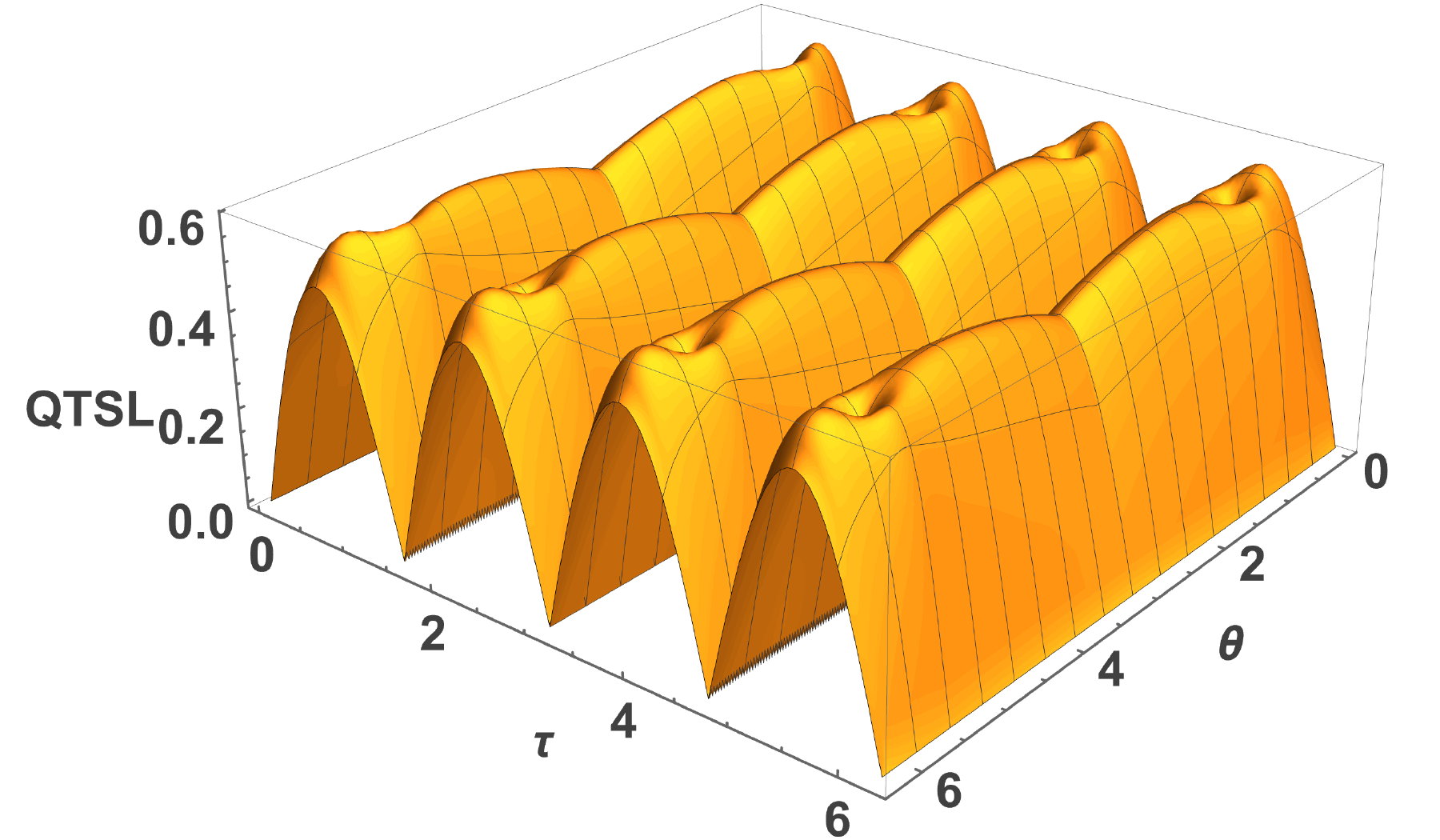}
\caption{
Dependence of the QTSL of order~$1$ on $\theta$ (ranging from $0$ to $2\pi$) and $\tau$ (ranging from $0$ to $2\pi$) for the C-maybe state defined in Eq.~\eqref{eq:cmaybe}.
}
\label{fig:QTSL_C_MAYBE}
\end{figure}

Finally, the explicit expression for $T_1^{\star}$ is
\begin{align}
T_1^{\star}(\tau,\theta)=\frac{L(\tau,\theta)}{\Lambda(\tau,\theta)}\,,
\end{align}
where
\begin{align}
\begin{split}
L(\tau,\theta)&\equiv 
\abs{\sin(2\tau)}\!\left(\abs{\sin(\theta)}+\frac{1}{1+e^2}
\sqrt{2(1-e^4)\cos(\theta)+\frac{1+e^4}{2}(3+2\cos(2\theta))+2e^2\sin^2(\theta)\cos(4\tau)}\,\right),\\
\Lambda(\tau,\theta)&\equiv 
\frac{1}{\tau}\!\left(\left(2N+(-1)^N\sin(2\tau)\right)\abs{\sin(\theta)}+
\frac{1+e^2+(1-e^2)\cos(\theta)}{2(1+e^2)}\,\mathfrak{E}\!\left[4\tau\,\Bigg|\,\left(\frac{2e\sin(\theta)}{1+e^2+(1-e^2)\cos(\theta)}\right)^2\right]\right),
\end{split}
\end{align}
and $N = \left\lfloor\frac{2\tau}{\pi}+\frac{1}{2}\right\rfloor$.
The resulting dependence of $T_1^{\star}$ on $\theta$ and $\tau$ is illustrated in Fig.~\ref{fig:QTSL_C_MAYBE}.

\subsection{Werner-like states}
\label{app:WernerExplicit}

We now consider the case where the system and memory are initialized in a mixture of a maximally-mixed state and an arbitrary rank-1 projector,
\begin{align}
    \rho_{sm}(\varphi) = \frac{1-\lambda}{4}\id + \lambda\, \dya{\psi(\varphi)}\,,~ \lambda \in [0,1]\,.
    \label{eq:WernerLike}
\end{align}
Perhaps the most well-known example of two-qubit states of this form is the family of Werner states~\cite{werner1989quantum}, where the projector $\dya{\psi(\varphi)}$ represents an entanglement given by
\begin{align}
\ket{\psi(\varphi)} = \cos(\varphi)\ket{a_1}_s  \otimes\ket{b_1}_m  + \sin(\varphi)\ket{a_2}_s  \otimes\ket{b_2}_m ,
    \label{eq:WernerSuperposition}
\end{align}
with $\{\ket{a_1}, \ket{a_2}\}$ and $\{\ket{b_1}, \ket{b_2}\}$ forming orthonormal bases of $\mathcal{H}_s$ and $\mathcal{H}_m$, respectively. It has been shown~\cite{luo2018typical,deng2009sufficient} that 
\begin{align} 
0\leq \lambda\leq \frac{1}{1+2\sin(2\varphi)}\,, 
\end{align} 
indicates that $\rho_{sm}(\varphi)$ is separable, whereas it becomes entangled when this condition is violated.

Combining the initial states of the work reservoir and bath taken from Eq.~\eqref{eq:InitialStateWBExample}, the full initial state is 
\begin{align}
    \rho_{\tot}(0) 
        = \frac{1-\lambda}{4}\id \otimes \dya{1}_w \otimes \rho_b^{th} 
        + \lambda \dya{\psi(\varphi)} \otimes \dya{1}_w \otimes \rho_b^{th}.
\end{align}
Under the dynamics generated by the Hamiltonians in Eqs.~\eqref{eq:BareHamiltonianExample} and \eqref{eq:InteractionHamiltonianExample}, the first of these two terms remains stationary.
Therefore, the time-evolved state is
\begin{align}
    \rho_{\tot}(t) 
        = U(t) \rho_{\tot}(0) U^\dagger(t)
        = \frac{1-\lambda}{4}\id \otimes \dya{1}_w \otimes \rho_b^{th} 
        + \lambda U(t)\left(\dya{\psi(\varphi)} \otimes \dya{1}_w \otimes \rho_b^{th}\right)U^\dagger(t).
\end{align}
When computing the time-averaged $p$-norm of Eq.~\eqref{eq:time-averaged-p-norm}, it is the derivative of $\rho(t)$ with respect to $t$ that enters and hence Eq.~\eqref{eq:time-averaged-p-norm} only depends on the second term,
\begin{align}
    \Lambda_s  ^{(p)}(\tau) 
        &= \frac{\lambda}{\tau}\int_0^\tau \norm{\partial_{t} \Tr_{m,w,b}\left\{U(t)\left[\dya{\psi(\varphi)} \otimes \dya{1}_w \otimes \rho_b^{th}\right]U^\dagger(t)\right\}}_{p}dt,
    \\
    \Lambda_m ^{(p)}(\tau) 
        &= \frac{\lambda}{\tau}\int_0^\tau \norm{\partial_{t} \Tr_{s,w,b}\left\{U(t)\left[\dya{\psi(\varphi)} \otimes \dya{1}_w \otimes \rho_b^{th}\right]U^\dagger(t)\right\}}_{p}dt.
\end{align}
Therefore, for both the system and memory it is guaranteed that $\Lambda_{s,m}^{(p)} \propto \lambda$.

We come to a similar result when computing the Schatten $p$-distance between the initial and final system-memory density operators as it appears in the numerator of Eq.~\eqref{eq:qsl-time}, the QSL time.
Since the mixed-state component of $\rho$ does not change, it does not contribute to the distance and we find that the distance is also proportional to $\lambda$,
\begin{align}
\!\!\!\!\!\!\!\!
\begin{split}
\ell_p(\mathcal{E}_s  (\rho_s  ), \rho_s  )
        &= \lambda\norm{\Tr_{w,b,m}\left\{U(\tau)\left[\dya{\psi(\varphi)} \otimes \dya{1}_w \otimes \rho_b^{th}\right]U^\dagger(\tau)\right\} - \Tr_{w,b,m}\left\{\dya{\psi(\varphi)} \otimes \dya{1}_w \otimes \rho_b^{th}\right\} }_p,\\
\ell_p(\mathcal{E}_m (\rho_m ), \rho_m )
        &= \lambda\norm{\Tr_{w,b,s}\left\{U(\tau)\left[\dya{\psi(\varphi)} \otimes \dya{1}_w \otimes \rho_b^{th}\right]U^\dagger(\tau)\right\} - \Tr_{w,b,s}\left\{\dya{\psi(\varphi)} \otimes \dya{1}_w \otimes \rho_b^{th}\right\} }_p\,.
\end{split}
\end{align}
With these results in mind, we expect that both the QSL time of Eq.~\eqref{eq:qsl-time} and the QTSL of Eq.~\eqref{eq:QTSL} will be completely independent of the mixing parameter $\lambda$ for Werner-like states. In the remainder of this appendix, we verify this result for the Werner state of Eq.~\eqref{eq:WernerLike} under the dynamics generated by the Hamiltonians in Eqs.~\eqref{eq:BareHamiltonianExample} by computing the QTSL explicitly.

\subsubsection{When QTSL is dependent on $\varphi$}
To explicitly compute the QTSL, it is necessary to choose some pair of bases used to define the superposition of Eq.~\eqref{eq:WernerSuperposition}.
Here we choose to have $\{\ket{a_1}_s\otimes\ket{b_1}_m,\ket{a_2}_s\otimes\ket{b_2}_m\}$ be the $Z_s\otimes X_m$ bases, so that at least one of the two bases does not coincide with the natural basis for the Hamiltonian.
Then, $\ket{\psi(\varphi)}$ becomes
\begin{align}
\ket{\psi(\varphi)}=\cos(\varphi)\ket{0}_s\otimes\ket{+}_m+\sin(\varphi)\ket{1}_s\otimes\ket{-}_m\,.
\end{align}

\begin{figure}[htp!]
\centering
\includegraphics[width=0.6\columnwidth]{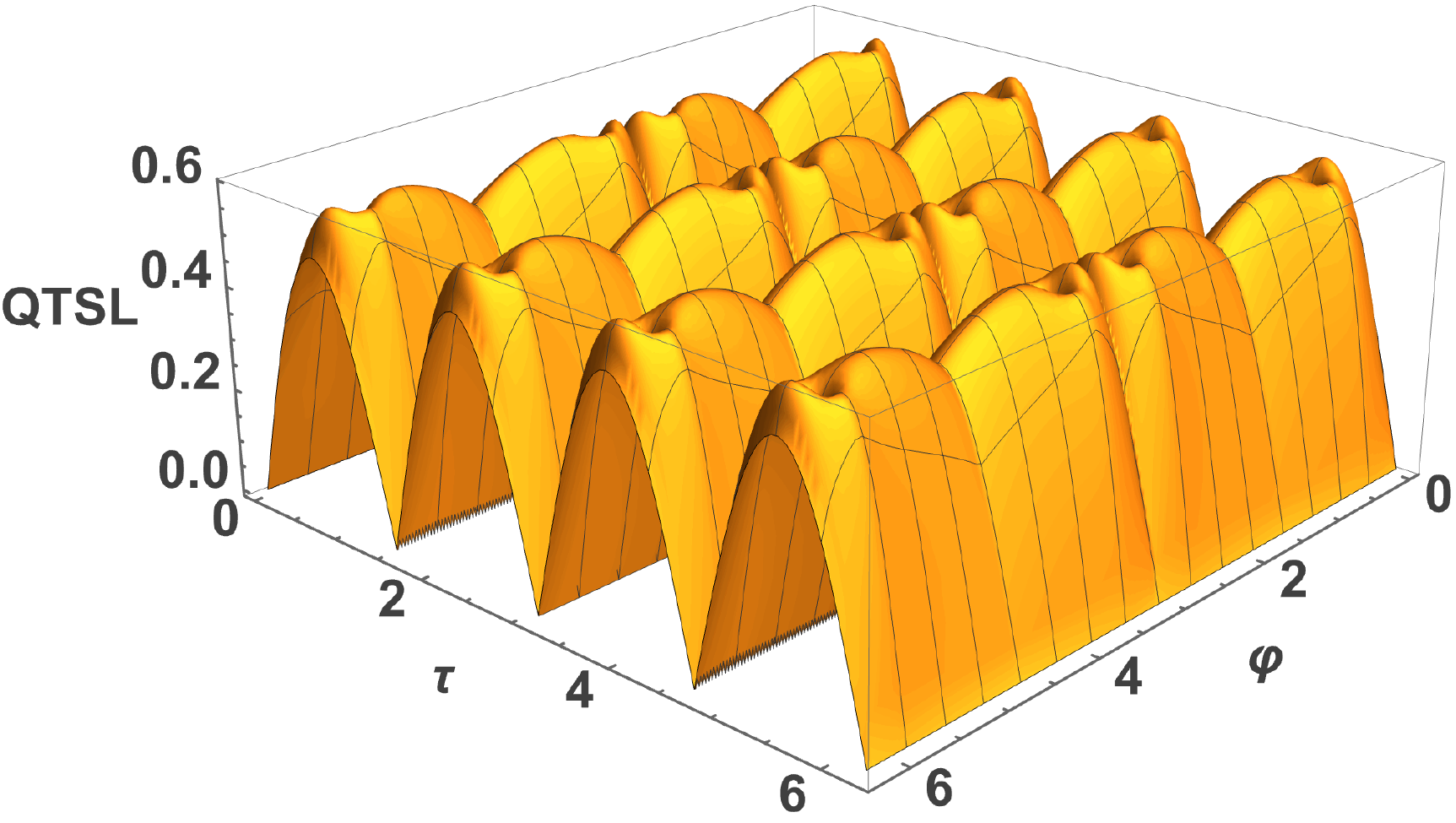}
\caption{
Dependence of the QTSL of order~$1$ on $\varphi$ (ranging from $0$ to $2\pi$) and $\tau$ (ranging from $0$ to $2\pi$) for the Werner-like state with $Z_s\otimes X_m$ basis.
}
\label{fig:QTSL_Werner_ZX}
\end{figure}

The initial and time-evolved system and memory states are
\begin{align}
    \rho_s&=\frac{1}{2}
    \begin{pmatrix}
        1+\lambda\cos(2\varphi)&0\\
        0&\frac{1}{2}(1-\lambda\cos(2\varphi))
    \end{pmatrix}\\
    \rho_m&=\frac{1}{2}
    \begin{pmatrix}
        1&\lambda\cos(2\varphi)\\
        \lambda\cos(2\varphi)&1
    \end{pmatrix}\\
    \mathcal{E}_s(\rho_s)&=\frac{1}{2}
    \begin{pmatrix}
      1+\lambda\cos(2\varphi)&-i\lambda \sin(2t)\sin(2\varphi)\\
      i\lambda \sin(2t)\sin(2\varphi)&1-\lambda\cos(2\varphi)
    \end{pmatrix}\\
    \mathcal{E}_m(\rho_m)&=\frac{1}{2}
    \begin{pmatrix}
       1&\lambda\frac{e^{-4it}(1+e^{2+4it})(\cos^2(\varphi)-e^{4it}\sin^2(\varphi))}{1+e^2}\\
       \lambda\frac{e^{-4it}(e^2+e^{4it})(e^{4it}\cos^2(\varphi)-\sin^2(\varphi))}{1+e^2}&1
    \end{pmatrix}\,.
\end{align}
Taking $p=1$, the distance between the initial and final states for the system and memory are
\begin{align}
 \begin{split}   \ell_1\left(\mathcal{E}_s(\rho_s),\rho_s\right)&=\lambda\abs{\sin(2\varphi)\sin(2\tau)}\\
\ell_1\left(\mathcal{E}_m(\rho_m),\rho_m\right)&=\frac{2\lambda}{1+e^2}\abs{\sin(2\tau)}\sqrt{\cos^4(\varphi)+e^4\sin^4(\varphi)+\frac{e^2\sin^2(2\varphi)}{2}\cos(4\tau)}\,.
 \end{split}
\end{align}
and the time-averaged $p=1$ norms are obtained as follows: for $\Lambda_s^{(1)}(\tau)$, from Eqs.~\eqref{eq:integralcos2t_1} and \eqref{eq:integralcos2t_2}, we have 
\begin{align}
    \Lambda_s^{(1)}(\tau)=\frac{1}{\tau}\int_{0}^{\tau}2\lambda \abs{\sin(2\varphi)\cos(2t)}dt=\frac{\lambda}{\tau}\left(2N+(-1)^N\sin(2\tau)\right)\abs{\sin(2\varphi)}\,,
\end{align}
where $N=\left\lfloor\frac{2\tau}{\pi}+\frac{1}{2}\right\rfloor$. For $\Lambda_m^{(1)}(\tau)$, from Eq.~\eqref{eq:EllipticFunction}, we have 
\begin{align}
\begin{split}
    \Lambda_m^{(1)}(\tau)&=\frac{1}{\tau}\int_{0}^{\tau}\frac{4\lambda}{1+e^2}\sqrt{\cos^4(\varphi)+e^4\sin^4(\varphi)+\frac{e^2\sin^2(2\varphi)}{2}\cos(8t)}dt\\
    &=\lambda\frac{\cos^2(\varphi)+e^2\sin^2(\varphi)}{\tau(1+e^2)}\mathfrak{E}\left[4\tau\,\Bigg|\,\left(\frac{e\sin(2\varphi)}{\cos^2(\varphi)+e^2\sin^2(\varphi)}\right)^2\right]\,.
\end{split}
\end{align}
Therefore, we can obtain the QTSL 
\begin{align}
    T_1^{\star}=\frac{L(\tau,\varphi)}{\Lambda(\tau,\varphi)}\,,
\end{align}
with 
\begin{align}
    \begin{split}
        L(\tau,\varphi)&\equiv\abs{\sin(2\tau)}\left(\abs{\sin(2\varphi)}+\frac{2}{1+e^2}\sqrt{\cos^4(\varphi)+e^4\sin^4(\varphi)+\frac{e^2\sin^2(2\varphi)}{2}\cos(4\tau)}\right)\\
        \Lambda(\tau,\varphi)&\equiv\frac{1}{\tau}\left(\left(2N+(-1)^N\sin(2\tau)\right)\abs{\sin(2\varphi)}+\frac{\cos^2(\varphi)+e^2\sin^2(\varphi)}{1+e^2}\mathfrak{E}\left[4\tau\,\Bigg|\,\left(\frac{e\sin(2\varphi)}{\cos^2(\varphi)+e^2\sin^2(\varphi)}\right)^2\right]\right)\,.
    \end{split}
\end{align}
The relationship between $T_1^{\star}$, $\varphi$ and $\tau$ is illustrated in Fig.~\ref{fig:QTSL_Werner_ZX}. 
Note that these results share many features of the results obtained in Fig.~\ref{fig:QTSL_C_MAYBE}. 

It is evident from the final expressions for $L$ and $\lambda$ that $T_1^{\star}$ does not depend on either $\lambda$.
This demonstrates that the presence or absence of entanglement between the system and memory does not necessarily have any impact on the QTSL.

\subsubsection{When QTSL is independent of $\varphi$}
Other bases show different behavior, depending on the relationship between the initial state and the preferred bases of the the Hamiltonian. When $\{\ket{a_1}\otimes\ket{b_1},\ket{a_2}\otimes\ket{b_2}\}$ are the $X_s\otimes X_m$ bases, the QTSL becomes independent of $\varphi$. 
In this case, $\ket{\psi(\varphi)}$ is given by
\begin{align} 
\ket{\psi(\varphi)} = \cos(\varphi)\ket{+}_s  \otimes\ket{+}_m  + \sin(\varphi) \ket{-}_s  \otimes\ket{-}_m \,. 
\end{align} 
Then, we have 
\begin{align}
\rho_s&= 
\begin{pmatrix} 
\cos^2\!\left(\frac{\theta}{2}\right)&0\\ 
0&\sin^2\!\left(\frac{\theta}{2}\right)
\end{pmatrix},\\
\rho_m&=\frac{1}{2}
\begin{pmatrix}
1+\sin(\theta)\cos(\theta)&\cos^2(\theta)\\
\cos^2(\theta)&1-\sin(\theta)\cos(\theta)
\end{pmatrix},\\
\mathcal{E}_s(\rho_s)&=\frac{1}{4}
\begin{pmatrix}
4\cos^2\!\left(\frac{\theta}{2}\right)&i\sin(2\theta)\sin(2\tau)\\
-i\sin(2\theta)\sin(2\tau)&4\sin^2\!\left(\frac{\theta}{2}\right)
\end{pmatrix},\\
\mathcal{E}_m(\rho_m)&=\frac{1}{2}
\begin{pmatrix}
1+\sin(\theta)\cos(\theta)&\frac{e^{-4i\tau}(1+e^{2+4i\tau})\cos(\theta)\left(1+e^{4i\tau}(-1+\cos(\theta))+\cos(\theta)\right)}{2(1+e^2)}\\
\frac{e^{-4i\tau}(e^2+e^{4i\tau})\cos(\theta)\left(-1+\cos(\theta)+e^{4i\tau}(1+\cos(\theta))\right)}{2(1+e^2)}&1-\sin(\theta)\cos(\theta)
\end{pmatrix}\,,
\end{align}
which leads to
\begin{align} 
\begin{split} 
\ell_1(\mathcal{E}_s(\rho_s),\rho_s)&=2\,\lambda\,\abs{\cos(2\varphi)}\sin^2(\tau)\propto\lambda\\ 
\ell_1(\mathcal{E}_m(\rho_m),\rho_m)&=\lambda\,\abs{\cos(2\varphi)\sin(2\tau)}\frac{\sqrt{1+e^4-2e^2\cos(4\tau)}}{e^2+1}\propto\lambda\,. 
\end{split} 
\end{align} 
Also, for $\Lambda_s^{(1)}(\tau)$, we have 
\begin{align} 
\Lambda_s^{(1)}(\tau)
=\frac{ 2\,\lambda\,\abs{\sin(\varphi)+\cos(\varphi)}\sqrt{1-\sin(2\varphi)}}{\tau}\int_0^{\tau}\abs{\sin(2t)}dt\,. 
\end{align} 
For $\frac{N\pi}{2}\leq\tau < \frac{(N+1)\pi}{2}$, we can write 
$N = \left\lfloor\frac{2\tau}{\pi}\right\rfloor$, and we obtain 
\begin{align} 
\int_{0}^{\tau}\abs{\sin(2t)}dt
=N+\frac{1-(-1)^N\cos(2\tau)}{2}\,. 
\end{align} 
Using this result, we find the time-averaged norms
\begin{align} 
\Lambda_s^{(1)}(\tau)
=\lambda\,\abs{\sin(\varphi)+\cos(\varphi)}\,\frac{\sqrt{1-\sin(2\varphi)}}{\tau}\left( 2N+1-(-1)^{N}\cos(2\tau)\right)\propto\lambda\,,
\end{align}
and
\begin{align}
\begin{split}
\Lambda_m^{(1)}(\tau)&=\frac{2\,\lambda\,\abs{\sin(\varphi)+\cos(\varphi)}\sqrt{1-\sin(2\varphi)}}{e^2+1}\int_0^{\tau}\sqrt{1+e^4-2e^2\cos(8t)}dt\,, 
\\
&=\lambda\,\abs{\sin(\varphi)+\cos(\varphi)}\,\frac{\sqrt{1-\sin(2\varphi)}}{\tau}\,\frac{e^2-1}{2(e^2+1)}\,\,\mathfrak{E}\left[4\tau\,\,\Bigg|\,\,-\left(\frac{2e}{e^2-1}\right)^2\right]\propto\lambda\,
\end{split}
\end{align} 
where we have again made use of Eq.~\eqref{eq:EllipticFunction} to re-express the integral in $\Lambda_m^{(1)}(\tau)$,
\begin{align} 
\int_0^{\tau}\sqrt{1+e^4-2e^2\cos(8t)}dt&=\frac{e^2-1}{4}\,\mathfrak{E}\left(4\tau\,\,\Bigg|\,\,-\left(\frac{2e}{e^2-1}\right)^2\right)\, .
\end{align} 
Since following ratio is independent of the angle $\varphi$, as
\begin{align} 
\frac{\abs{\cos(2\varphi)}}{\abs{\sin(\varphi) + \cos(\varphi)}\sqrt{1 - \sin(2\varphi)}}
=\frac{\sqrt{1+\sin(2\varphi)}\sqrt{1-\sin(2\varphi)}}{\sqrt{\left(\sin(\varphi)+\cos(\varphi)\right)^2}\sqrt{1-\sin(2\varphi)}}= \sqrt{\frac{1+\sin(2\varphi)}{1 + 2\sin(\varphi)\cos(\varphi)}}
= 1\,,
\end{align}
$T_1^{\star}$ is equal to the ratio
\begin{align} 
T_1^{\star}(\tau)=\frac{L(\tau)}{\Lambda(\tau)}\,, 
\end{align} 
as defined in Eq.~\eqref{eq:QTSL}, with $L(\tau)$ and $\Lambda(\tau)$ being
\begin{align} 
\begin{split} 
L(\tau)&\equiv 2\sin^2(\tau)+\abs{\sin(2\tau)}\frac{\sqrt{1+e^4-2e^2\cos(4\tau)}}{e^2+1}\\ 
\Lambda(\tau) &\equiv \frac{1}{\tau}\,\left(2N+1-(-1)^{N}\cos(2\tau)+\frac{e^2-1}{2(e^2+1)}\,\mathfrak{E}\left[4\tau\,\Bigg|\,-\left(\frac{2e}{e^2-1}\right)^2\right]\right)\,, 
\end{split} 
\end{align} 
with $N = \left\lfloor2\tau / \pi\right\rfloor$.

\begin{figure}[htp!] 
\centering 
\includegraphics[width=0.6\columnwidth]{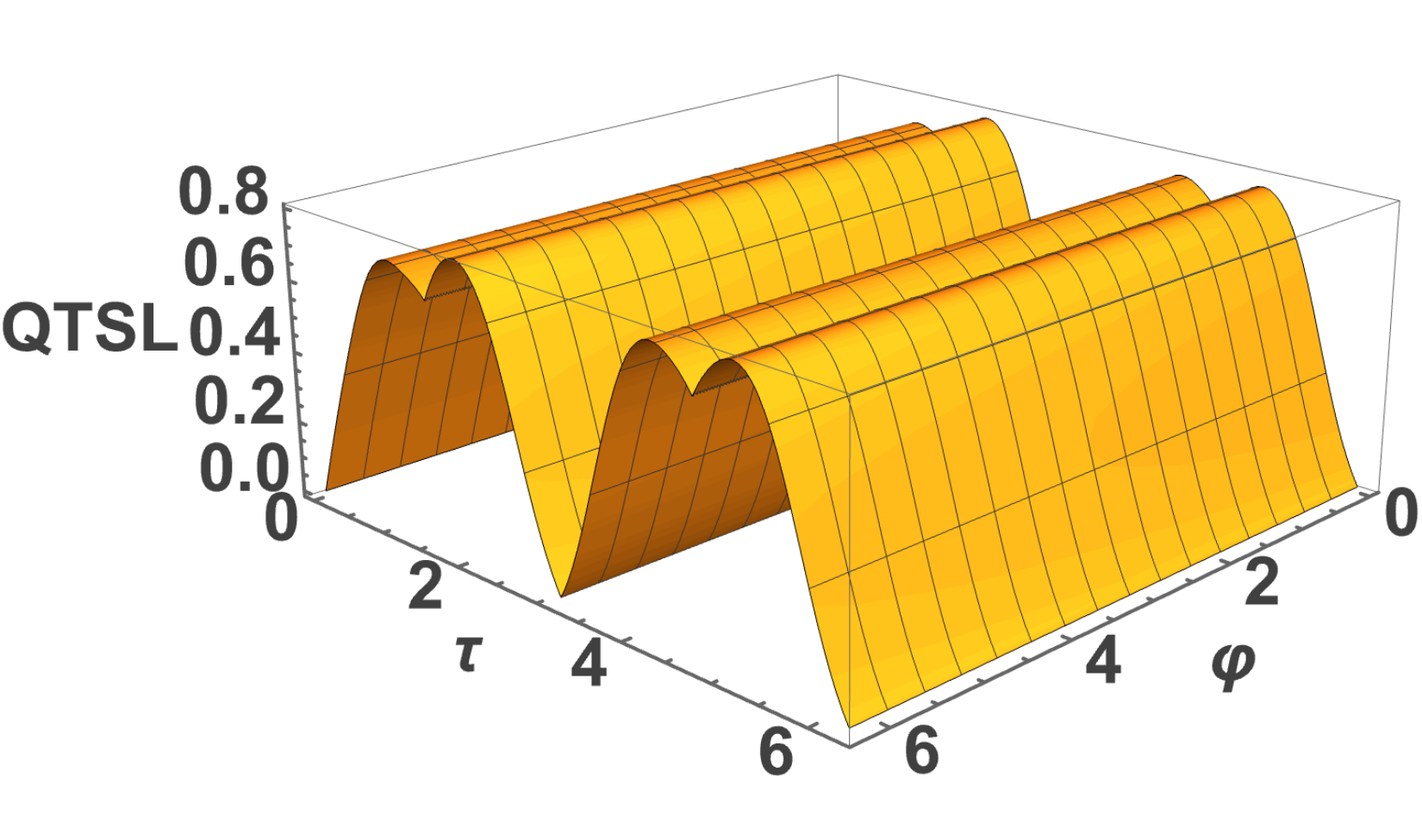} 
\caption{Dependence of the QTSL of order $p=1$ on $\varphi$ (ranging from $0$ to $2\pi$) and $\tau$ (ranging from $0$ to $2\pi$) for the Werner-like state with $X_s\otimes X_m$ basis.} 
\label{fig:QTSL_Werner_3D} 
\end{figure}

The relationship between $T_1^{\star}$, $\varphi$, and $\tau$ is shown in Fig.~\ref{fig:QTSL_Werner_3D}. 
It is clear from the plot that the relationship between $T_1^{\star}$ and $\tau$ in this case differs significantly from previous two examples, indicating that the behavior of the QTSL can be highly dependent on the relationship between the initial state and the Hamiltonian.

\subsubsection{General statement}

As we have seen, the QTSL with Werner-like initial states may or may not depend on the angle $\varphi$, depending on the specific bases selected to define the superposition.
The independence of the QTSL on the mixing parameter $\lambda$, however, does not.
Further, this independence is not unique to the example Hamiltonian we have considered thus far but is in fact generally true.

To show this, consider generic initial states for the work reservoir and bath combined with a Werner-like state for the system and memory,
\begin{align}
    \rho_{\tot}(0) = \left(\frac{1-\lambda}{4}\id + \lambda \dya{\psi(\varphi)}\right) \otimes \rho_w \otimes \rho_b^{th}
    .
\end{align}
The first component of this state will remain stationary under the dynamics generated by $H_{\tot}$ if,
\begin{align}
    \left[H_s + H_m + H_w + H_b + H_{sm} + H_{sw} + H_{sb} + H_{mb}, \id_s \otimes \id_m \otimes \rho_w \otimes \rho_b^{th} \right] = 0
    .
    \label{eq:id-commutator-req}
\end{align}
Since the initial system and memory states are maximally mixed, it is clear that
\begin{align}
    \left[H_s + H_m + H_{sm}, \id_s \otimes \id_m \otimes \rho_w \otimes \rho_b^{th}\right] = 0
    .
\end{align}
By the definition of a thermal state, we also have that
\begin{align}
    \left[H_b, \id_s \otimes \id_m \otimes \rho_w \otimes \rho_b^{th}\right] = 0
    .
\end{align}
For the two Hamiltonians involving the work reservoir $H_{w}, H_{sw}$, we may use a form satisfying Eq.~\eqref{eq:compatible} to show that their contribution to Eq.~\eqref{eq:id-commutator-req} also vanishes,
\begin{align}
    \left[H_w + H_{sw}, \id_s \otimes \id_m \otimes \rho_w \otimes \rho_b^{th}\right] = 0
    .
\end{align}

What remains is to show that $H_{sb}$ and $H_{mb}$ commute with this state. 
Take $H_{sb}$ (the same arguments work for $H_{mb}$), and consider decomposing it into bi-partite operators
\begin{align}
    H_{sb} = \sum_{j=1}^N A_j \otimes B_j
    ,
\end{align}
with $\{A_j\}$ a collection of distinct operators acting on the system and $\{B_j\}$ a collection of not necessarily distinct operators acting on the bath. 
Then the commutator,
\begin{align}
    \left[H_{sb}, \id_s \otimes \rho_b^{th}\right] 
     = \sum_{j=1}^N A_j \otimes \left[B_j, \rho_b^{th}\right]
     ,
\end{align}
vanishes only if 
\begin{align}
    \left[B_j, \rho_b^{th}\right] = 0~(\forall j)\,.
\end{align}
That is, we must require that the bath remains stationary in its equilibrium thermal state under the action of $H_{sb}$. Similar arguments produce the same result for $H_{mb}$.

All together, we have that 
\begin{align}
    \left[H_{\tot}, \id_s \otimes \id_m \otimes \rho_w \otimes \rho_b^{th}\right] = 0
    ,
\end{align}
whenever we require that the bath remains quasistatic in its equilibrium thermal state and that the dynamics generated by $H_{\tot}$ ensure that the work source acts as a catalyst with initial state $\rho_w$ as discussed in Sec.~\ref{sec:unitaryAndhamiltonian}.
Therefore, it will always be the case that the QTSL for Werner-like states depends only on the identity of the projector involved and not on the mixing parameter.

\twocolumngrid

\bibliography{ref.bib}

\end{document}